\documentclass[letterpaper, 10 pt, conference]{ieeeconf}  

\IEEEoverridecommandlockouts                              

\usepackage{graphics} 
\usepackage{epsfig} 
\usepackage{mathptmx} 
\usepackage{amsmath} 
\usepackage{amssymb}  
\usepackage{svg}
\usepackage{url}
\usepackage{amsmath}

\usepackage{breakurl}
\usepackage[breaklinks]{hyperref}
\usepackage[absolute]{textpos}
\title{\LARGE \bf
A Data-Driven Integrated Framework for Fast-Charging Facility Planning using Multi-Period Bi-Objective Optimization
}

\author{Mingjia He$^{1}$, Panchamy Krishnakumari$^{2}$, Ding Luo $^{3}$ and Jiaqi Chen $^{4}$
\thanks{$^{1}$Mingjia He is with the Faculty of Civil Engineering and Geosciences,
        Delft University of Technology, 2628 CN, Delft, The Netherlands
        {\tt\small M.He-3@student.tudelft.nl}}%
\thanks{$^{2}$ Panchamy Krishnakumari is with the Faculty of Civil Engineering and Geosciences,
         Delft University of Technology, 2628 CN Delft, The Netherlands
        {\tt\small P.K.Krishnakumari@tudelft.nl}}%
\thanks{$^{3}$ Ding Luo is with Shell plc, 1031 HW Amsterdam, The Netherlands
        {\tt\small Ding.Luo@shell.com}}%
\thanks{$^{4}$ Jiaqi Chen is with Shell plc, 1031 HW Amsterdam, The Netherlands
        {\tt\small Jiaqi.Chen@shell.com}}%
}

\begin{document}
\begin{textblock}{18}(2,1)
\noindent\textcolor{blue}{In the Proceedings of the 2023 IEEE 26th International Conference on Intelligent Transportation Systems (ITSC)} 
\end{textblock}

\maketitle
\thispagestyle{empty}
\pagestyle{empty}

\begin{abstract}

With the electrification in freight transportation, the availability of fast-charging facilities becomes essential to facilitate en-route charging for freight electric vehicles. Most studies focus on planning charging facilities based on mathematical modeling and hypothetical scenarios. This study aims to develop a data-driven integrated framework for fast-charging facility planning. By leveraging the highway traffic data, we extracted, analyzed, and compared spatial and temporal flow patterns of general traffic and freight traffic. Furthermore, graph theory-based network evaluation methods are employed to identify traffic nodes within the highway network that play a significant role in accommodating charging infrastructure. A candidate selection method is proposed to obtain potential deployment locations for charging stations and to-go chargers. Based on this, we present a multi-period bi-objective optimization model to provide optimal solutions for the placement of charging facilities, with the objectives of minimizing investment cost and maximizing demand coverage. The case study on the Amsterdam highway network shows how existing traffic data can be used to generate more realistic charging demand scenarios and how it can be integrated and evaluated within the optimization framework for facility planning. The study also shows that the proposed model can leverage the potential of early investment in improving the charging demand coverage.

\end{abstract}

\section{INTRODUCTION}

Transportation has become one of the major contributing sectors to emissions, accounting for approximately one-quarter of all greenhouse gas emissions in Europe \cite{EE2019}. The Netherlands is ambitious to achieve zero-emission road traffic by 2050.  With incentivizing policies and tax-related measures, the Netherlands has become one of the leading electric transport players in the world.  In freight transport, Netherlands' government plans to raise the market share of clean heavy-duty vehicles to reach 30\% by 2030 \cite{G2021}. Considering the ambition of zero-emission policy and the current developing  trend, the market for electric freight vehicles will grow continuously and thus requires the construction of new charging infrastructure.\\
There are two main types of charging solutions: alternating current (AC) slow charging and direct current (DC) fast charging. AC charging is mainly served for destination charging at workplaces or residences, as it requires more time to load. An AC slow charger may take 6–8 hours to recharge the vehicle battery to full state, while a DC fast charger can recharge up to 80\% within about 30 minutes \cite{Yang2021}. The high efficiency of DC fast chargers is attributed to the higher voltage and direct flow of DC current into the battery without conversion. This characteristic makes DC charging a promising solution for long-distance travel \cite{Yang2021}. It allows en-route charging to ease driving anxiety and driving range restrictions. The current charging infrastructure is insufficient to support the future growth of EVs. In particular, the existing charging infrastructure lacks enough fast chargers \cite{Xia2022}. In 2019, there are more than 200 fast-charging stations in the Netherlands. Researchers expect significant growth in the number of fast-charge points for electric cars over the coming years, to a maximum of 8,000 by 2025 \cite{N2017}. Amsterdam, The Hague, Rotterdam, Utrecht, and Brabantstad have been designated as the focus areas to develop charging infrastructure since 2009 \cite{N2019}.\\
To promote electrification in freight transport, the goal of this study is to propose fast-charging infrastructure planning strategies for the en-route charging of commercial freight vehicles along the highway. This study builds a planning framework consisting of data fusion, network evaluation, candidate location selection, and an optimization model for planning. A multi-period bi-objective optimization model is constructed to find the optimal locations and scales of fast charging facilities considering the investment and charging-demand coverage. The case study is conducted based on the Amsterdam highway network. The study provides evidence for long-term charging facility investment and supports the electrification of intercity logistics.

\section{LITERATURE REVIEW}

With the increasing market share of electric vehicles in road transportation, extensive research has investigated the charging infrastructure-planning problem. Based on the way to represent charging demand, research approaches can be categorized into the node-based model, flow-based model, and trajectory-based model \cite{Boujelben2021,Lin2019}. In the node-based model, it is assumed that the charging demand is generated at the nodes in the network \cite{He2016,Jung2014}. The flow-based model uses a set of origin–destination trips and allows charging demand to be served during journeys \cite{Wu2017}. The trajectory-based model considers the travel pattern of electric vehicles \cite{Yang2017, Tu2016} and might incorporate the individual charging decision and route scheduling \cite{Liu2019}. Optimization models would be established after obtaining charging demand. Many researchers considered multiple objectives for various benefits of different stakeholders.
Yang et al. \cite{Yang2021} established bi-objective programming models for charging demand assignment, fast charging station operation, and power line expansion, with objectives to maximize charging service profit and minimize total charging time.
Bian et al. \cite{Bian2022} proposed the charging station configuration model from the perspective of users considering traffic congestion and signal-lights waiting time. To find the Pareto optimal solution set, the simulated annealing particle swarm optimization algorithm was used with objectives of minimum investment cost, maximum profitability, and minimum time-consuming cost. 
Liu et at. \cite{Liu2022} established the bi-level planning model for electric vehicle charging stations and used the firefly algorithm to find solutions. The upper model optimized the location and capacity of charging stations with the objective of maximizing the annual profit. The lower model optimized individual electric vehicle charging plans to achieve minimum charging cost.
Wang et al. \cite{Wang2022} proposed an optimization model for the planning of slow-charging piles and fast-charging piles, incorporating the impact of road traffic conditions on the user’s charging additional cost. To efficiently find the Pareto solution sets, the NSGA-II algorithm was improved by modifying the initial population generation and crossover operator. The algorithm was proved to have better performance in terms of searchability and global convergence.\\
The development of charging infrastructure is likely to take several years in practice. Considering the dynamic charging demand and limited investment, it is difficult to deploy all the charging stations within one-step planning\cite{Kadri2020}. Some researchers have suggested that sing-stage optimization could lack the capacity to deal with long-term charging demand dynamics \cite{Li2022}. Charging infrastructure planning can be formulated as a sequential decision-making process, enabling the construction strategies to be changed according to charging demand \cite{Mehrjerdi2020,Vashisth2022}.
Meng el at. \cite{Meng2020} selected candidate charging station sites based on social limitations and proposed a sequential expansion-downsizing strategy for station construction. The proposed method provided flexible construction plans to balance the increase and decrease of charging demand. The objective was to minimize the total social cost by incorporating drivers’ cost and construction investment. 
Kadri et al. \cite{Kadri2020} used a multi-stage stochastic integer programming approach to address uncertainties in both EV-trip numbers within the road network and EV flows within trips. Scenario trees were used to approximate the evolution of the stochastic process over time, and the benders decomposition approach was extended to find optimality. Compared to the deterministic model, the proposed stochastic one provided a significantly greater coverage of charging demand.
Previous studies have provided in-depth insights into the charging station location problem. A majority of charging facility deployment strategies are based solely on mathematical modeling and implemented in hypothetical scenarios. As real-world data becomes more accessible and informative, more research is needed to develop data-driven planning methods capturing valuable information from a variety of sources (traffic flow, point-of-interest (POI) information, network configuration). In addition, the charging infrastructure layout should fit into the structure of the road network, reflecting the characteristic of the network. Yet, research into potential charging station locations has rarely considered network evaluation. Furthermore, many studies have applied multi-period planning and multi-objective planning in recent years, but few have combined these two aspects into a model that takes into account both demand dynamics and benefit trade-offs. To this end, the contribution of this study has three folds: 1) to leverage the information of freight traffic data and POI data into the charging facility planning process; 2) to propose a comprehensive selection process of candidate locations for charging facilities incorporating charging demand, network structure, interests of service providers, and construction flexibility; 3) to develop a multi-period bi-objective optimization model considering the charging demand dynamics over years and the trade-offs between total cost and demand coverage.

\section{METHODOLOGY}

\begin{figure}[thpb]
    \centering
    \includegraphics[width=7cm]{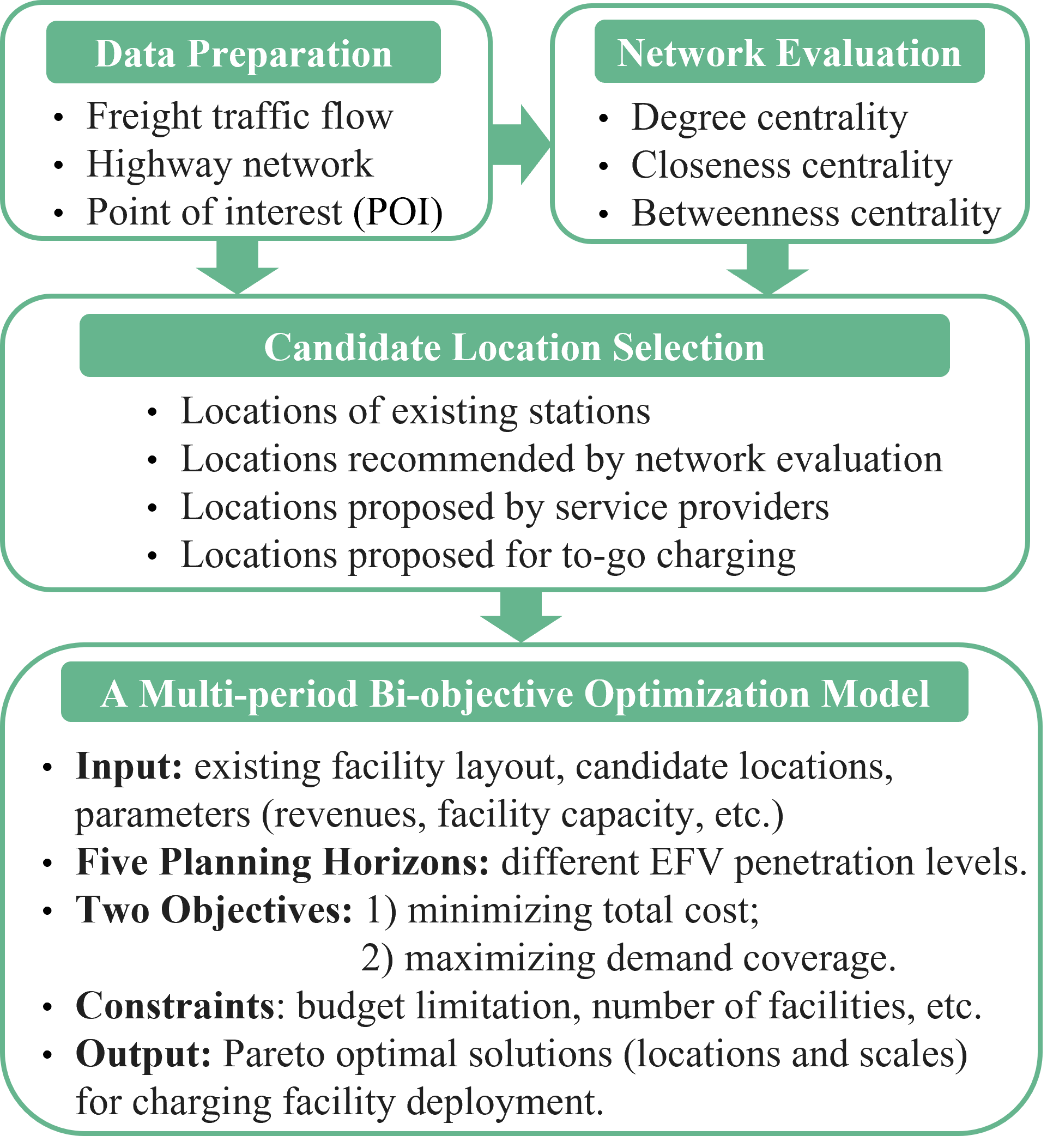}
    \caption{Methodological framework}
    \label{fig:framework}
\end{figure}

This study will model the charging facility planning problem and provide insight into how charging facility providers can make construction plans for the future of freight transportation electrification. The proposed methodological framework shown in Figure \ref{fig:framework} consists of four parts: data preparation, network evaluation, candidate location selection, and charging location optimization. Data preparation and network evaluation leverage the valuable information of data and knowledge of graph theory into the planning process. Using indicators of centrality, the rankings of nodes within the highway network can be determined. We will identify the nodes that play a more significant role in the network by evaluating the connections among nodes. Moreover, a clear procedure for selecting candidate locations is established. For integrated planning, the mathematical model considered multi-period optimization with two objectives minimizing total cost and maximizing demand coverage. 

 \subsection{Data Preparation}
The proposed framework will be implemented on the real highway network in the Netherlands. The datasets required in this study include freight traffic flow data, highway network data, and POI data. Traffic data on highways can be obtained from the data website, NDW \cite{NDW2022}. We extracted one-week traffic data from the date 2022-06-27 to 2022-07-03. The information includes the date, time period, route ID, flow, speed, and vehicle class. Freight traffic flow data can be used to determine freight charging demand. As a percentage of the total traffic flow, the market penetration rate is used to determine the amount of traffic to be charged.  The datasets of the highway road network and POI data can be obtained from Open Street Map \cite{OSM2022}. POI data would provide information on the category of locations and geographical coordinates. 
\subsection{Network Evaluation}
The highway network can be defined as an undirected graph $G=(N,A)$, where $N$ represents the set of nodes (highway junctions) and $A=\{(i,j),i,j \in N, i \neq j\}$ represents the set of arcs (roads). Network evaluation would answer the question of how important a node is in the highway network. To evaluate the role of nodes, centrality indicators are calculated including degree centrality, closeness centrality, and betweenness centrality.
Degree centrality ($DC_i$) measures the number of connected nodes (Equation (1)). Nodes with a high degree score have higher connectiveness. Closeness centrality ($CC_i$) measures the average inverse distance to all other nodes, reflecting a node’s closeness to others (Equation (2)). Nodes with a high closeness score have a shorter total distance to all other nodes. Betweenness centrality ($BC_i$) represents the degree to which nodes stand between each other. It involves calculating the shortest paths between all pairs of nodes in the network (Equation (3)).
\begin{equation}
    DC_i= \dfrac{D} {N-1} 
    \label{dc}
\end{equation}
\vspace{-6pt}
\begin{equation}
    CC_i=\dfrac {1} {\sum_{i,j\in G,i \neq j} d_{i,j}} 
    \label{cc}
\end{equation}
\vspace{-6pt}
\begin{equation}
    BC_i=\sum_{i,j,v\in G,i \neq j\neq v} {\dfrac {\sigma_{j,v}(i)}{\sigma_{j,v}}}
    \label{bc}
\end{equation}
Where $N$ is the total number of nodes; $d_{i,j}$ represents the shortest path length between node $i$ and node $j$; $G$ is the vertices set in the network; $\sigma_{j,v}$ is the total number of shortest paths from node $j$ and node $v$; $\sigma_{j,v}(i)$ is the number of those paths passing through node $i$.

\subsection{Candidate Location Selection}

The charging facility planning model considers deploying charging facilities in the candidate locations rather than all possible locations. There are four types of candidate locations for the deployment of charging facilities: 1) those with existing facilities; 2) those recommended by network evaluation results; 3) those selected by service providers; and 4) those for to-go charging. \\
Existing facilities are included first on the list as the existing charging stations can be expanded. In the second type of candidate locations, the graph theory will determine the highway nodes of importance, and candidate locations will be selected around these nodes. Furthermore, this study includes locations chosen by service providers based on business considerations. In addition to building charging stations, the last type of candidate location considers that fast chargers can be deployed at supermarkets (instead of charging stations) to provide high-efficiency charging. For the selection process, those POIs with the labels `fuel station', `truck stop', and `parking area' are considered for existing facilities, while those labeled `supermarket' are for to-go charging. It should be noted that only POIs that are less than 500 meters from the highway are considered to serve enroute charging demand. 

\subsection{Multi-period Bi-objective 0ptimization Model}
A multi-stage optimization model is proposed with the objectives of minimizing the total cost and maximizing the coverage number of freight vehicles. Considering the development of transport electrification, it is assumed that the proportion of electric freight vehicles increases over the years. The notation for the optimization model is presented in Table \ref{tab:notation}.

\begin{table}[h]
    \caption{Notation}
    \label{tab:notation}
    \centering

    \begin{tabular}{p{0.5cm}p{7.5cm}}
    \hline
        \centering{Variable} & \quad \quad \quad {Description} \\
        \hline
          \multicolumn{2}{l}{Parameters}\\
          $K$ & Set of planning horizons.\\
          $I$ & Set of candidate locations for charging stations.\\
          $J$ & Set of candidate locations for to-go charging piles.\\
         $C_k$ & The total cost of the planning horizon $k$.\\
         $D_k$ & The demand coverage of the planning horizon $k$.\\
         $\gamma$ & The weight in the objective of demand coverage.\\
         $c^{x^{k-1}_i,x^k_i}_s$ & The cost of station i from the state $x^{k-1}_i$ to $x^k_i$. \\
         $c_t$ & The cost of installing one fast charging pile near the supermarket.\\
         $d^k_i$ & The demand coverage of charging station i in horizon $k$.\\
         $d^k_j$ & The demand coverage of to-go charging facilities $j$ in horizon $k$.\\
         $q^k_i$ & The freight flow can be covered by facility $i$ in horizon $k$.\\
         $p^k$ & Market penetration rate of electric freight vehicles in horizon $k$.\\
         $b^k$ & The maximum investment in horizon $k$.\\
         $cap_l$  &  The capacity of a charging station with the scale $l$.\\
         $cap_t$  &  The capacity of a to-go charging pile.\\
         $Nmin^k$  &  The minimum number of charging stations in horizon $k$.\\
         $Nmax^k$  &  The maximum number of charging stations in horizon $k$.\\
         $Mmin^k$  & The minimum number of locations for to-go chargers in horizon $k$. \\
         $Mmax^k$  & The maximum number of  locations for to-go chargers in horizon $k$. \\
        $dist_{i,j}$  &  The distance between station $i$ and station $j$. \\
        $dist_{min}$   &  The minimum distance between two charging stations. \\
        $s$  &  The maximum construction scale of charging stations.\\
        $n$  &  The maximum number of fast-charging plies near the supermarket.\\
        \multicolumn{2}{l}{Decision variables}\\
        $\eta^k_i$  &  Binary variable: whether a charging station is deployed at the location $i$ in horizon $k$. \\
        $\eta^k_j$  &  Binary variable: whether to-go chargers are deployed at the location $j$ in horizon $k$.\\
        $x^k_i$ & The construction scale of station $i$ in horizon $k$.\\
        $y^k_j$  &  The number of installed fast-charging piles near supermarket $j$ in horizon $k$.\\         
    \hline  
    \end{tabular}

\end{table}

\subsubsection{Planning horizons}
This model considers five planning horizons, representing different stages of development, with EFV penetration rates of 20\%, 40\%, 60\%, 80\%, and 100\%. The initial planning period uses the current charging facility layout. Starting from the second planning period, each subsequent period builds upon the layout of the previous periods. This means that the results of one horizon serve as the input for the optimization model of the next horizon.
 
\subsubsection{Model formulation}
The first objective is to minimize the total cost in Equation (4), considering that the service provider would control the project investment and reduce it as much as possible. As indicated by the previous research on charging facility planning, the total cost could be an influential factor in the scale of the planning project (e.g. the number and size of charging stations). The construction cost in each horizon consists of the cost of charging stations and the cost of to-go charging at supermarkets (in Equation (5)).

$$
    min \quad Z_1(k)=  C_k 
    \eqno{(4)}
$$
\vspace{-8pt}
$$
    C_k=\sum_{i=1}^{I} c^{x^{k-1}_i,x^k_i}_s +\sum_{j=1}^{J} (y^{k}_i-y^{k-1}_i)c_t, \quad k\in K
    \eqno{(5)}
$$
The second objective in Equation (6) is maximizing the coverage of charging demand. The parameter  $\gamma_k $ determines whether the next-horizon planning is included in the current planning objectives. In Equations (7) and (8), the demand coverage is determined by electric freight flow coverage and facility capacity. The electric freight flow coverage can be calculated by penetration rate $p^k$ multiplying the average freight flow at the nearest starting point of highway segments. The facility capacity is determined by the construction scale of charging stations and to-go charging facilities.

$$
    min \quad Z_2(k)= -( D_k + \gamma_k  D_{k+1})
    \eqno{(6)}
$$
\vspace{-8pt}
$$
    D_k = \sum_{i=1}^{I} min(q^k_ip^k,cap_l x^k_i)
    +\sum_{j=1}^{J} min(q^k_jp^k,cap_t y^k_j),$$$$ 
    \quad  i\in I, j\in J, k \in K \eqno{(7)}
$$
\vspace{-8pt}
$$
    \gamma_k = 
    \begin{cases}
    1 \quad \text{k=1,2,3,4}\\
    0 \quad \text{k=5}
    \end{cases}
    \eqno{(8)}
$$
\vspace{-8pt}
\\
\textit{Subject to constraints:}

$$
    \sum_{i=1}^{I} c^{x^{k-1}_i,x^k_i}_s +\sum_{j=1}^{J} (y^{k}_j-y^{k-1}_j)c_t < b^k, \quad k \in K \eqno{(9)}
$$
\vspace{-5pt}
$$
   \eta^k_i\eta^k_jdist_{i,j}<dist\_min, \quad i,j\in I,k \in K
   \eqno{(10)}
$$
\vspace{-8pt}
$$
    x^k_i \leq x^{k+1}_i,\quad i\in I, k \in K
    \eqno{(11)}
$$
\vspace{-8pt}
$$
    y^k_j \leq y^{k+1}_j,\quad j\in J, k \in K
    \eqno{(12)}
$$
\vspace{-8pt}
$$
    Nmin^k<\sum_{i=1}^{I}\eta^k_i<Nmax^k , \quad k \in K
    \eqno{(13)}
$$
\vspace{-8pt}
$$
    Mmin^k<\sum_{j=1}^{J}\eta^k_j<Mmax^k, \quad k \in K
    \eqno{(14)}
$$
\vspace{-8pt}
$$
   \eta^k_i \in (0,1),\quad i\in I, k \in K
   \eqno{(15)}
$$
\vspace{-8pt}
$$
   0 \leq x^k_i\leq s,\quad i\in I, k \in K
   \eqno{(16)}
$$
\vspace{-8pt}
$$
    0 \leq y^k_j\leq n,\quad j\in J, k \in K
    \eqno{(17)}
$$

The limitations for investment are set for each planning horizon. Constraints (9) ensure that the cost in each horizon can not exceed the pre-set value. Constraints (10) indicate that the distance between two stations should be larger than the minimum distance threshold. Constraints (11) and (12) ensure that the scale of charging stations and the number of to-go chargers can not decrease with development, as it is considered that the charging facilities constructed in the previous horizons will remain in the subsequent horizons. For the first horizon ($k=1$), the planning is based on the initial (existing) facility layout ($k=0$). Constraints (13) and (14) set restrictions on the number of facilities. 
Constraints (15) represents the binary decision variables on whether to build facilities in candidate locations. Constraints (16) and (17) define two sets of integer variables for the scale of charging facilities, namely, scales of charging stations $x^k_i$ and scales of to-go charging facilities $y^k_i$.

\subsubsection{Solution Algorithm}
\begin{figure}[!h]
    \centering
    \includegraphics[width=8cm]{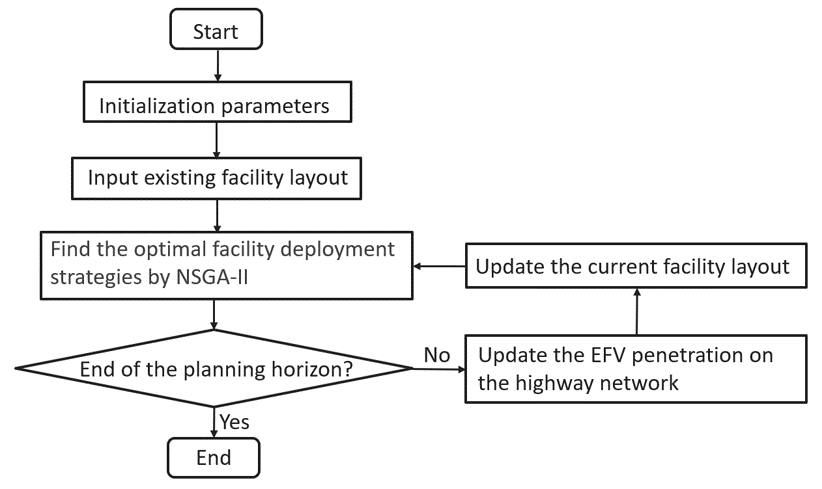}
    \caption{The procedure of optimization}
    \label{fig:horizon}
\end{figure}

we apply the non-dominated sorting genetic algorithm II (NSGA-II) to solve the bi-objective optimization problem with multiple horizons. 
Figure \ref{fig:horizon} shows how NSGA-II solves the proposed multi-period bi-objective charging facility planning model. In each horizon, NSGA-II will produce Pareto optimal solutions. These solutions will be compared to select one for the implementation and to be used to update the facility layout preparing for the next-period optimization. 

\section{CASE STUDY}
\subsection{Travel Pattern Analysis}
\begin{figure}[!h]
    \centering
    \includegraphics[width=9cm]{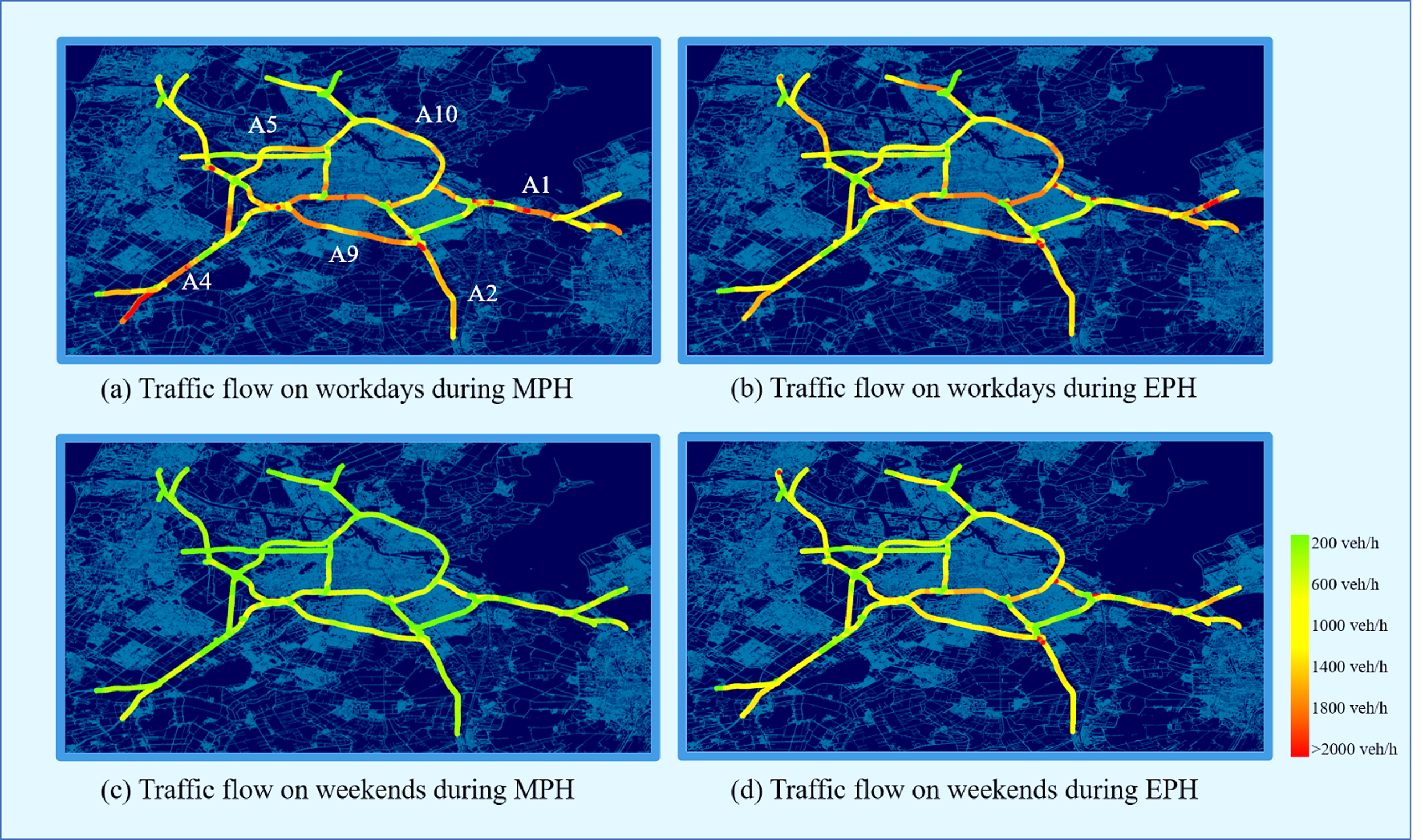}
    \caption{The distribution of overall traffic flow}
    \label{fig:traffic flow}
\end{figure}

\begin{figure}[!h]
    \centering
    \includegraphics[width=9cm]{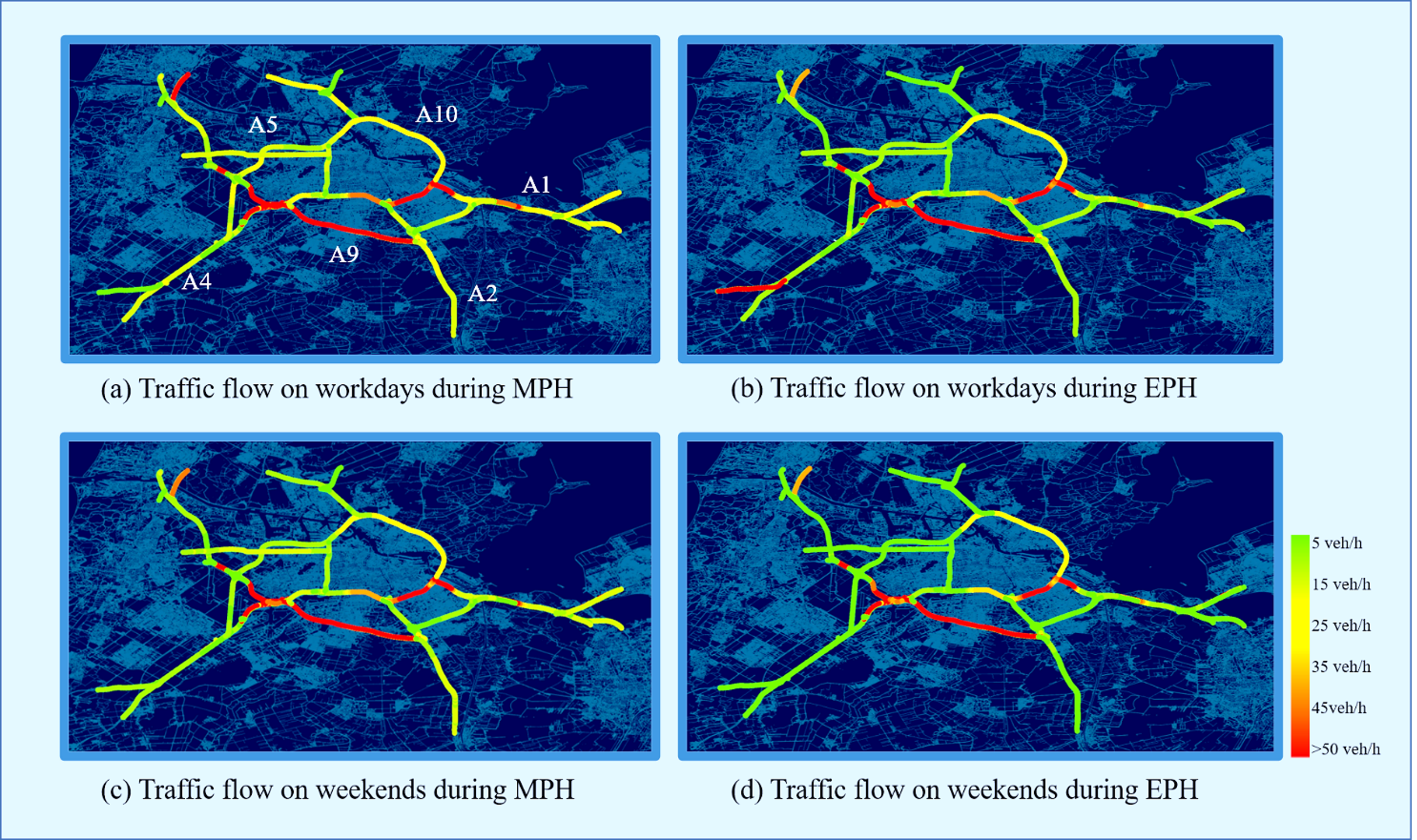}
    \caption{The distribution of truck traffic flow}
    \label{fig:truck flow}
\end{figure}

Figure \ref{fig:traffic flow} (a)-(d) depict traffic flow during morning and evening peak hours on both workdays and weekends. Comparing Figure \ref{fig:traffic flow} (a)(b) to Figure \ref{fig:traffic flow} (c)(d), it is evident that overall traffic demand was higher on workdays than weekends, in both morning and evening peak periods. During workday mornings, the northern part of the study area, including A4, A9, and A2, exhibited increased traffic demand. In the evenings, demand decreased, particularly in the north, while heavy traffic persisted on A9 and A1. On weekends (Figure \ref{fig:traffic flow} (c)), the road network experienced low traffic demand, with an average volume below 1400 vehicles per hour in the morning. However, during evening peak hours, there was an increase in traffic volume on A9, A4, and A1. The temporal and spatial distributions of truck demand (in Figure \ref{fig:truck flow} (a)-(d)) were relatively different from the overall traffic on the highway. Temporal patterns of truck volume were similar on workdays and weekends, morning and evening. Spatially, certain road segments (e.g., A9, A1/A10 intersection, southwest A10) consistently experienced high truck travel demand on both workdays and weekends. Suburban areas along A4 and A1 also had notable traffic demand.

\subsection{Candidate Location Selection}
To select the candidate locations, POI information is filtered according to the built-environment category. POIs with the label `fuel station', `parking area', `truck stop', and `supermarket' remained. It is noted that only the locations that lie in the 500-meter buffering of the highway would be used as candidates. Next, nodes in the highway network are evaluated. Table \ref{tab:evaluation} shows the information of the top 10 ranked candidates. The indicators, degree centrality (DC), closeness centrality (CC), and betweenness centrality (BC) are calculated and normalized. The score is the average of indicators. Node 17 is ranked the top with the highest values for all indicators, followed by Node 11, 5, and 3.

\begin{table}[!ht]
    \centering
    \caption{The numerical results of network evaluation}
    \label{tab:evaluation}
    \begin{tabular}{llllll}
    \hline
        Rank & Score & DC & CC & BC & ID \\ \hline
        1 & 1.673  & 1.000    & 0.310 & 0.363 & 17  \\ 
        2 & 1.609  & 1.000    & 0.267 & 0.342 & 11  \\ 
        3 & 1.489  & 1.000    & 0.287 & 0.202 & 5   \\ 
        4 & 1.446  & 1.000    & 0.265 & 0.181 & 3  \\ 
        5 & 1.250  & 0.667    & 0.277 & 0.306 & 20  \\ 
        6 & 1.209  & 0.667    & 0.292 & 0.249 & 18  \\ 
        7 & 1.188  & 0.667    & 0.295 & 0.226 & 4  \\ 
        8 & 1.164  & 0.667    & 0.248 & 0.249 & 25   \\ 
        9 & 1.136  & 0.667    & 0.267 & 0.202 & 24  \\ 
        10 & 1.130  & 0.667    & 0.225 & 0.239 & 12   \\ \hline
    \end{tabular}

\end{table}

\begin{figure}[!h]
    \centering
    \includegraphics[height=4cm]{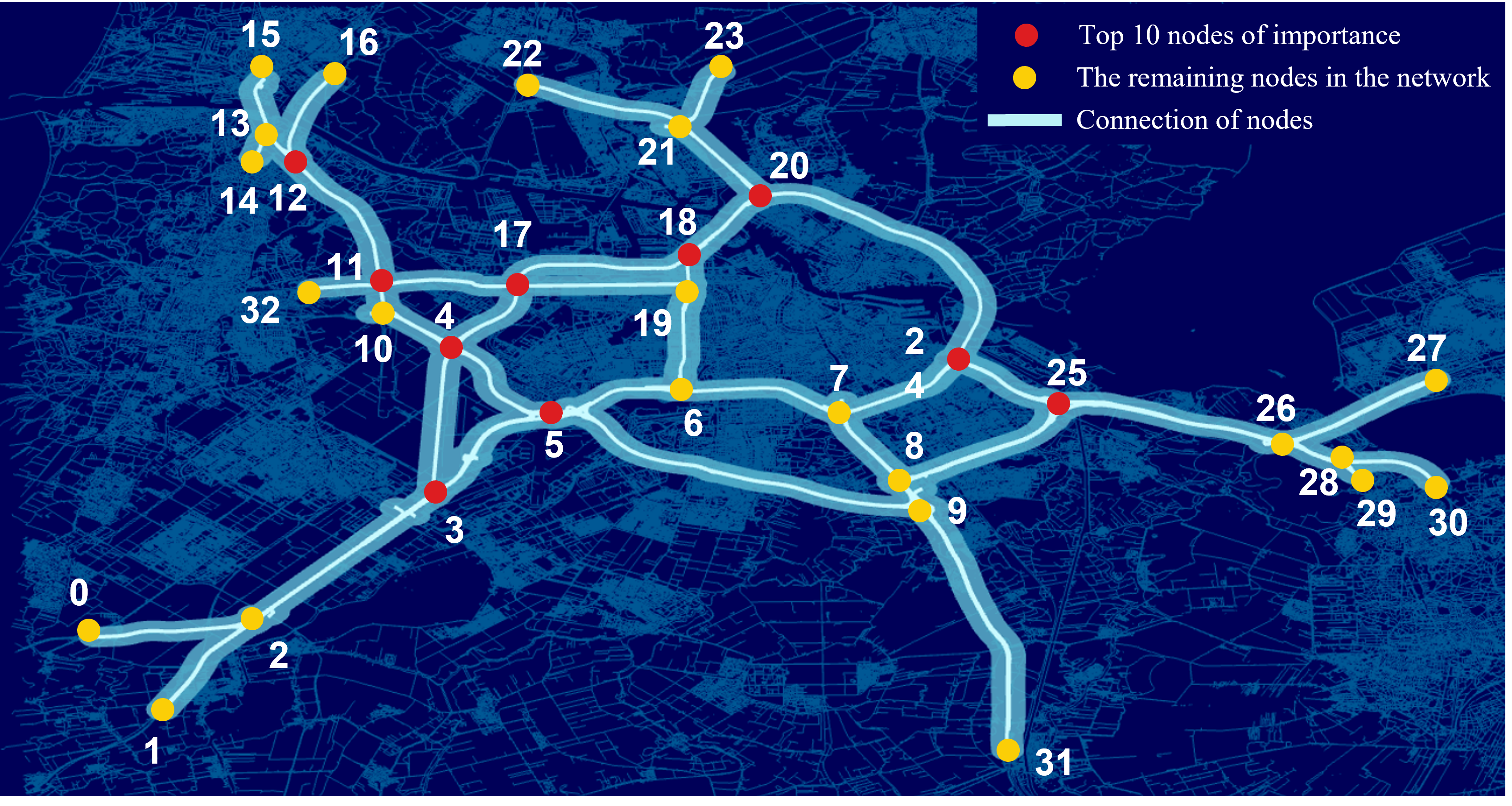}
    \caption{The distribution of evaluated nodes}
    \label{fig:network}
\end{figure}

\begin{figure}[!h]
    \centering
    \includegraphics[height=4cm]{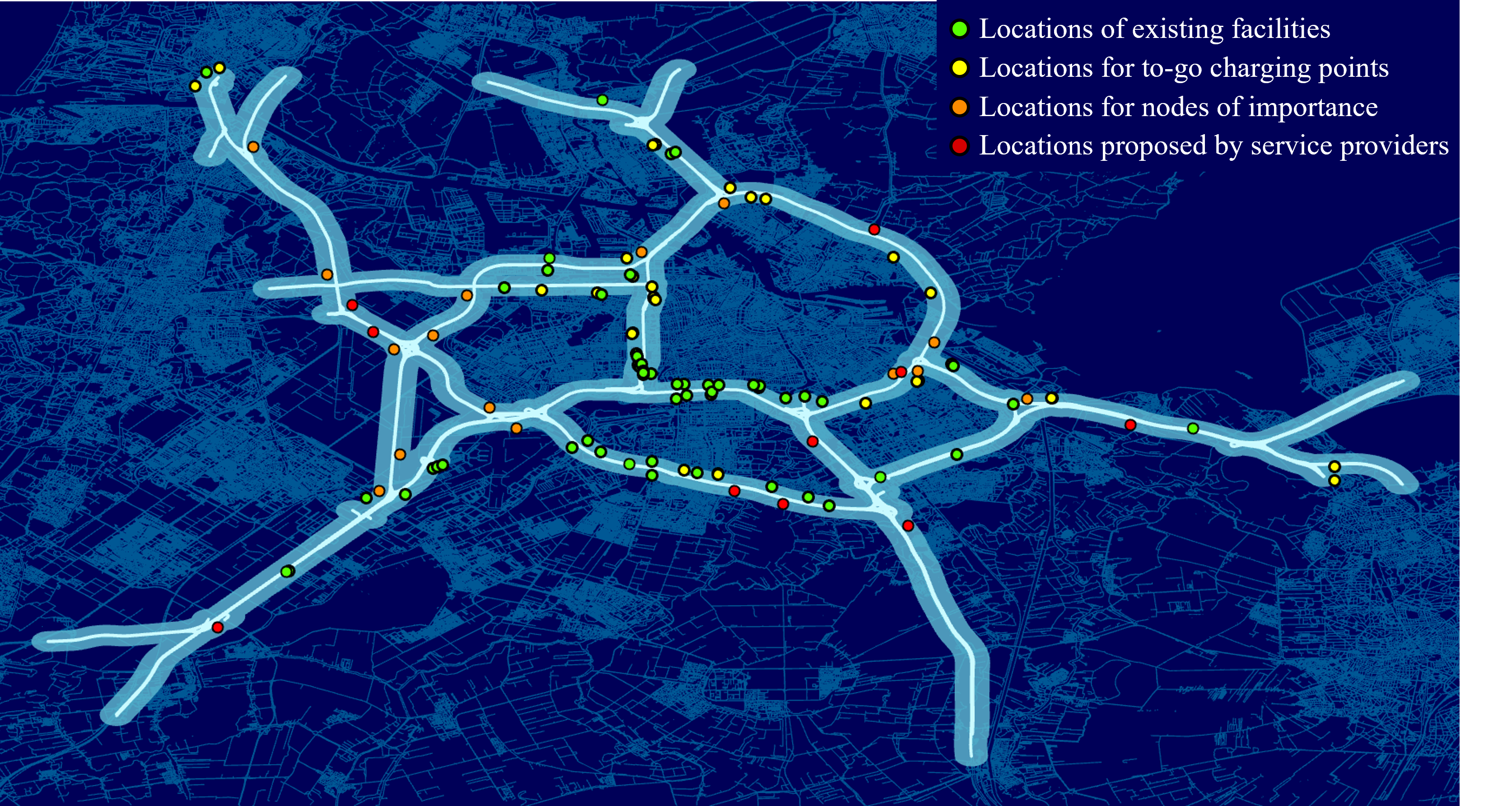}
    \caption{The distribution of evaluated nodes and candidate locations}
    \label{fig:candidate}
\end{figure}

Figure \ref{fig:network} shows the distribution of nodes in the highway network. The nodes in red color represented the nodes with the top 10 rankings, which play a more important role in this highway network. These nodes were the highway junctions that were more connected with other junctions and were more likely to influence other junctions in the network. As shown in Figure \ref{fig:candidate}, 119 candidate locations are selected in total, with 84 candidates for charging station deployment and 35 candidates for to-go charging points installment. The candidates selected by POI data accounted for the largest proportion 63.4\%. According to the network evaluation,  15 candidate locations were added near the top 10 nodes in the network, as indicated by the orange points. 

\subsection{Integrated Charging Facility Planning}
The optimization model will find the (near) optimal solutions for locations and construction scales of charging facilities. For charging stations, we define five construction scales for charging stations, extremely-small scale stations, small-scale stations, medium-scale stations, large-scale stations, and extremely-large-scale stations, represented by $x^k_i = (1,2,3,4,5)$. $x^k_i = 0$ means there is no station constructed at the location $i$. For to-go chargers at supermarkets, we optimize the number of charging piles to install ($y^k_i$). The optimization model has parameters in terms of investment, capacity, construction scale, distance, etc. The settings of parameters are presented in Table \ref{tab:value} in Appendix A. In setting these parameters, we have taken into account the parameter settings in previous research \cite{Li2021, Wu2021} and have tuned the parameters based on our case study. For NSGA-II algorithm, we set the population size to be 500, the iterations number to be 300, the crossover probability to be 0.9, and the mutation probability to be 0.1. 

\subsection{Multi-period Facility Planning}
During multistage planning, the initial results of the infrastructure planning can certainly affect subsequent planning stages. As bi-objective optimization can have more than one optimal solution in each horizon, called Pareto optimal solutions, one solution should be selected from the Pareto set for the next-horizon planning. To evaluate these impacts of solution selection, two scenarios are defined following different solution selection rules in each horizon planning. In Scenario 1, the solution with maximum demand coverage is selected for next-horizon planning. In Scenario 2, the solution with (the nearest) median demand coverage is selected. To evaluate the performance of looking ahead policy of our model, we define Scenario 3 in which the planning of each horizon only considers the demand in the current horizon, instead of the potential changing in the nearest future. In Scenario 3, the objective on the demand coverage only considers the effects on the current horizon, and the solution with maximum demand coverage was selected for next-horizon planning.\\
By comparing Scenario 1 and Scenario 2, we can observe the impact of solution selection on the final deployment layout. And Scenario 2 and Scenario 3 can show whether planning one step ahead can benefit long-term planning. As shown in Table \ref{tab:s1s2}, Scenario 1 has a total cost of 6.250 million euros (MER), and the demand coverage can reach 369 freight vehicles per hour in the last planning horizon. Scenario 2 saves 61\% of total cost compared to Scenario 1, the covered demand decreases by 29\%. Therefore, selecting maximum demand coverage in every horizon could obtain solutions with higher demand coverage, while it is noted that the cost-efficiency of investment could be smaller. When looking into the planning horizons, it can be found that Scenario 1 tends to construct new facilities as many as possible reaching the upper limit of the maximum number of facilities. The investigation of to-go chargers at the early stage indicates the advantage of flexible chargers at supermarkets for capturing charging demand with a relatively lower initial investment. The number of facilities in Scenario 2 is above the half level of that in Scenario 1.\\
Compared to Scenario 1, Scenario 3 neglects the growth of charging demand and tends to invest less in the first four horizons. Although in each horizon, Scenario 3 selects the solution with the largest demand coverage, it covers 94\% charging demand compared to that in Scenario 1. In the final horizon, Scenario 1 constructs more charging stations than Scenario 3 and has the same number of to-go chargers.

\begin{table}[!ht]
    \centering
    \caption{The optimization results of Scenarios}
    \label{tab:s1s2}
    \begin{tabular}{llll}
    \hline
         & Horizon 1 & Horizon 2 & Horizon 3 \\ \hline
        Scenario 1 & ~ & ~ & ~ \\ 
        Construction cost (MEUR) & 1.450  & 1.500  & 1.000\\ 
        Demand coverage (veh/h) & 98  & 204  & 256 \\ 
        Number of stations & 9 & 12 & 12  \\ 
        Number of to-go chargers & 25 & 25 & 25  \\ 
        Scenario 2 & ~ & ~ & ~  \\ 
        Construction cost (MEUR) & 0.03 & 0.508 & 0.508\\ 
        Demand coverage (veh/h) & 37  & 95  & 157 \\ 
        Number of stations & 7 & 8 & 9 \\ 
        Number of to-go chargers & 15 & 19 & 23 \\ 
        Scenario 3 & ~ & ~ & ~   \\ 
        Construction cost (MEUR) & 1.142  & 1.458  & 0.700 \\ 
        Demand coverage (veh/h) & 98  & 196  & 256  \\ 
        Number of stations & 9 & 11 & 12 \\ 
        Number of to-go chargers & 21 & 25 & 25\\ \hline
        
         &  Horizon 4 & Horizon 5 & Total\\ \hline
        Scenario 1 & ~ & ~ & ~  \\ 
        Construction cost (MEUR) &  1.300  & 1.000  & 6.250(100\%)\\ 
        Demand coverage (veh/h) &  316  & 369  & 369(100\%)\\ 
        Number of stations & 13 & 15 & 15  \\ 
        Number of to-go chargers & 25 & 25 & 25 \\ 
        Scenario 2 & ~ & ~ & ~  \\ 
        Construction cost (MEUR) & 0.502 & 0.9 & 2.448(39\%)\\ 
        Demand coverage (veh/h) & 203  & 262  & 262(71\%)\\ 
        Number of stations & 10 & 11 & 11\\ 
        Number of to-go chargers & 24 & 24 & 24\\ 
        Scenario 3 & ~ & ~ & ~   \\ 
        Construction cost (MEUR) & 0.950  & 1.100  & 5.350(86\%)\\ 
        Demand coverage (veh/h) & 304  & 346  & 346(94\%)\\ 
        Number of stations & 12 & 12 & 12\\ 
        Number of to-go chargers & 25 & 25 & 25\\ \hline
        
    \end{tabular}
\end{table}

\begin{figure}[!h]
    \centering
    \includegraphics[height=4.6cm]{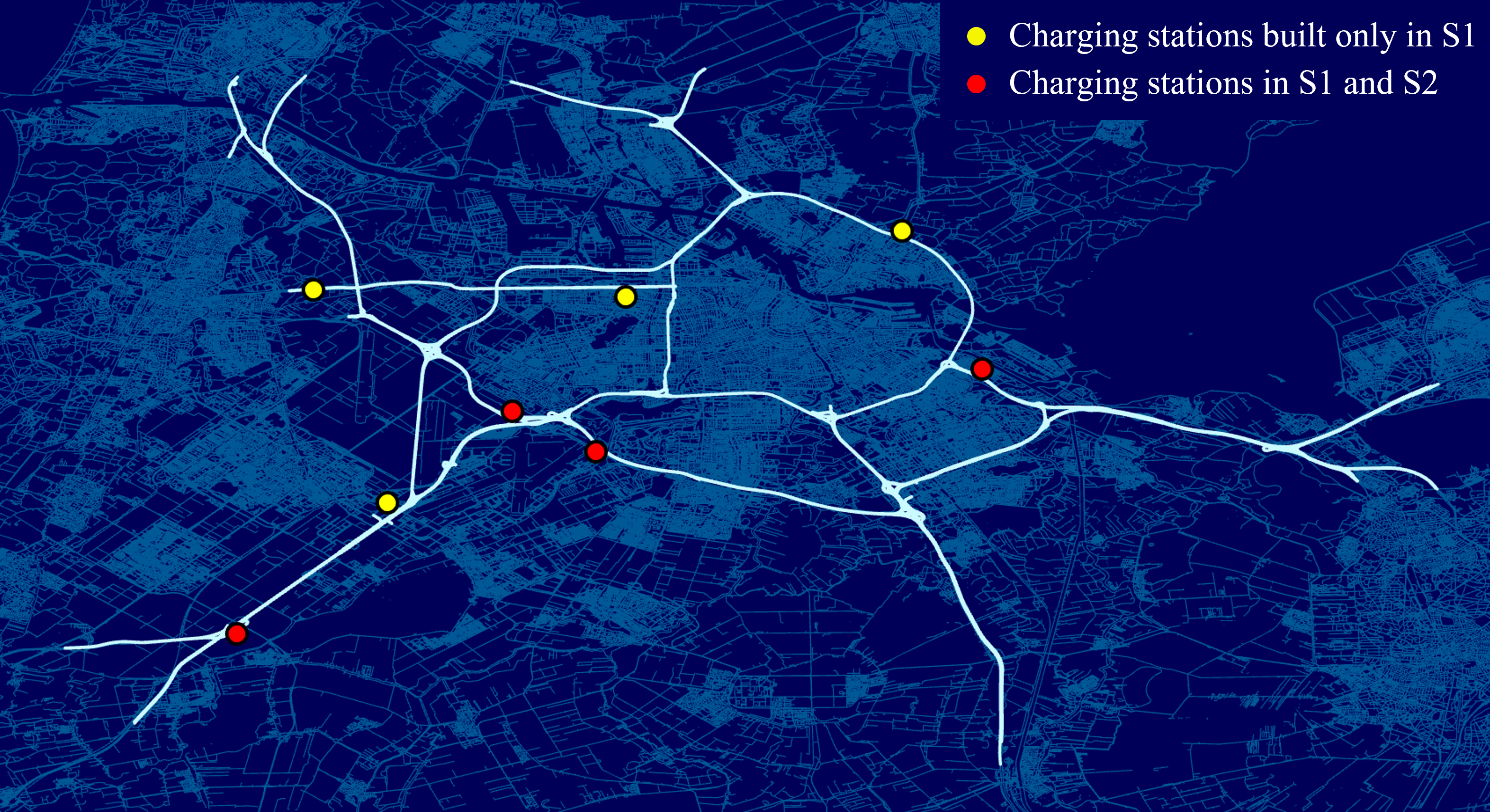}
    \caption{The distribution of charging stations in Scenario 1 and Scenario 2}
    \label{fig:stations}
\end{figure}

Figure \ref{fig:stations} shows the distribution of charging facility deployment plans of Scenario 1 and Scenario 2 respectively. In order to display the locations of newly constructed stations, existing stations are not shown. The yellow points represent the constructed charging stations only in Scenario 1. The red points represent the charging stations constructed both in Scenario 1 and Scenario 2. It should be noted that the locations for charging stations in Scenario 2 are also selected in Scenario 1. The overlap of scenarios may help facility planners to identify the locations that are more cost-efficient in the configuration of optimal deployment strategies. The spatial distribution of these stations is in line with the highway segments with high freight traffic flow in Figure \ref{fig:truck flow}.

\subsection{Sensitivity Analysis on Investment Limitation }
\begin{figure}[!h]
         \centering
         \includegraphics[height=5.5cm]{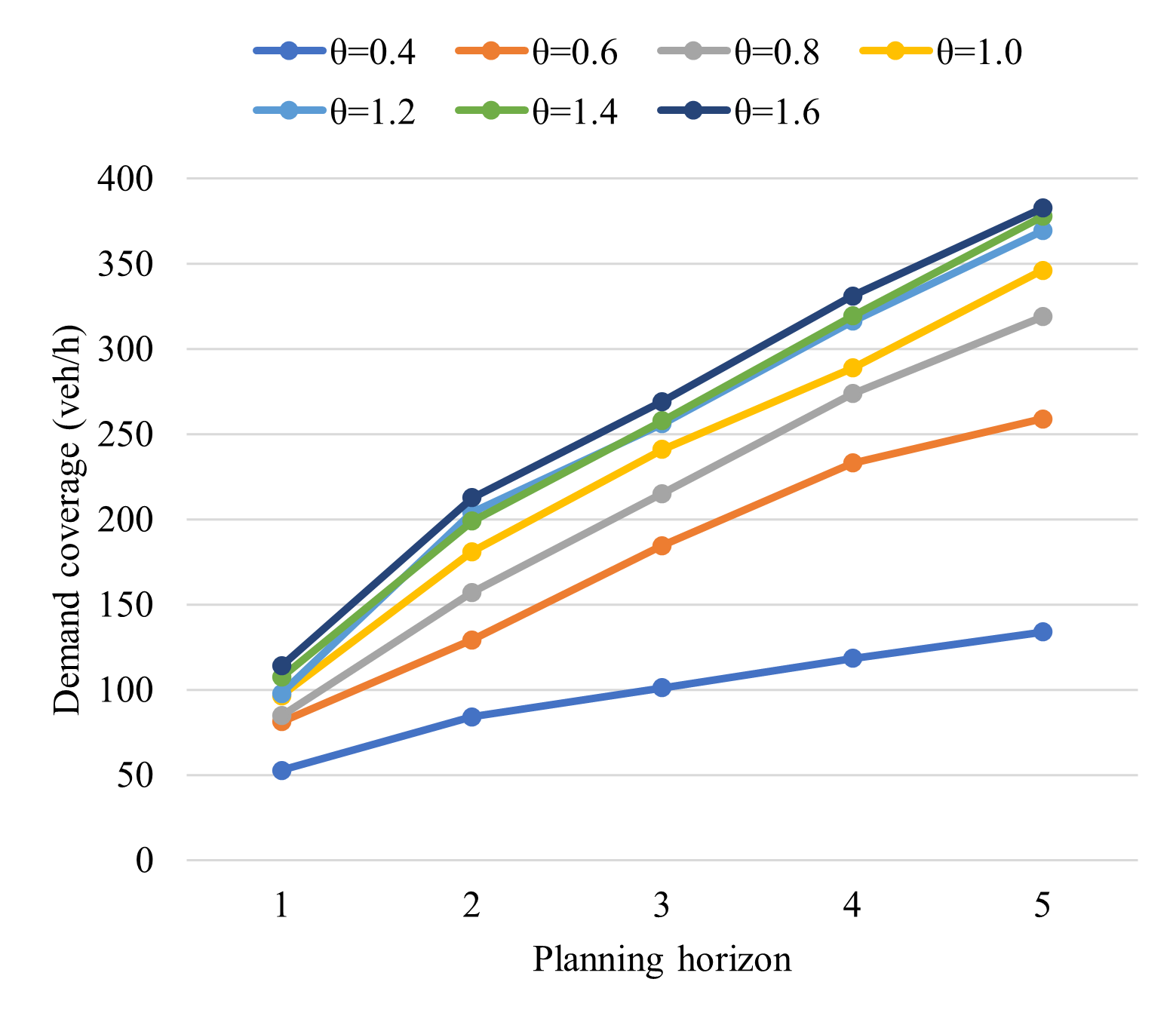}
         \caption{The demand coverage in each planning horizon}
         \label{fig:b1}
\end{figure}

\begin{figure}[!h]
         \centering
         \includegraphics[height=5.5cm]{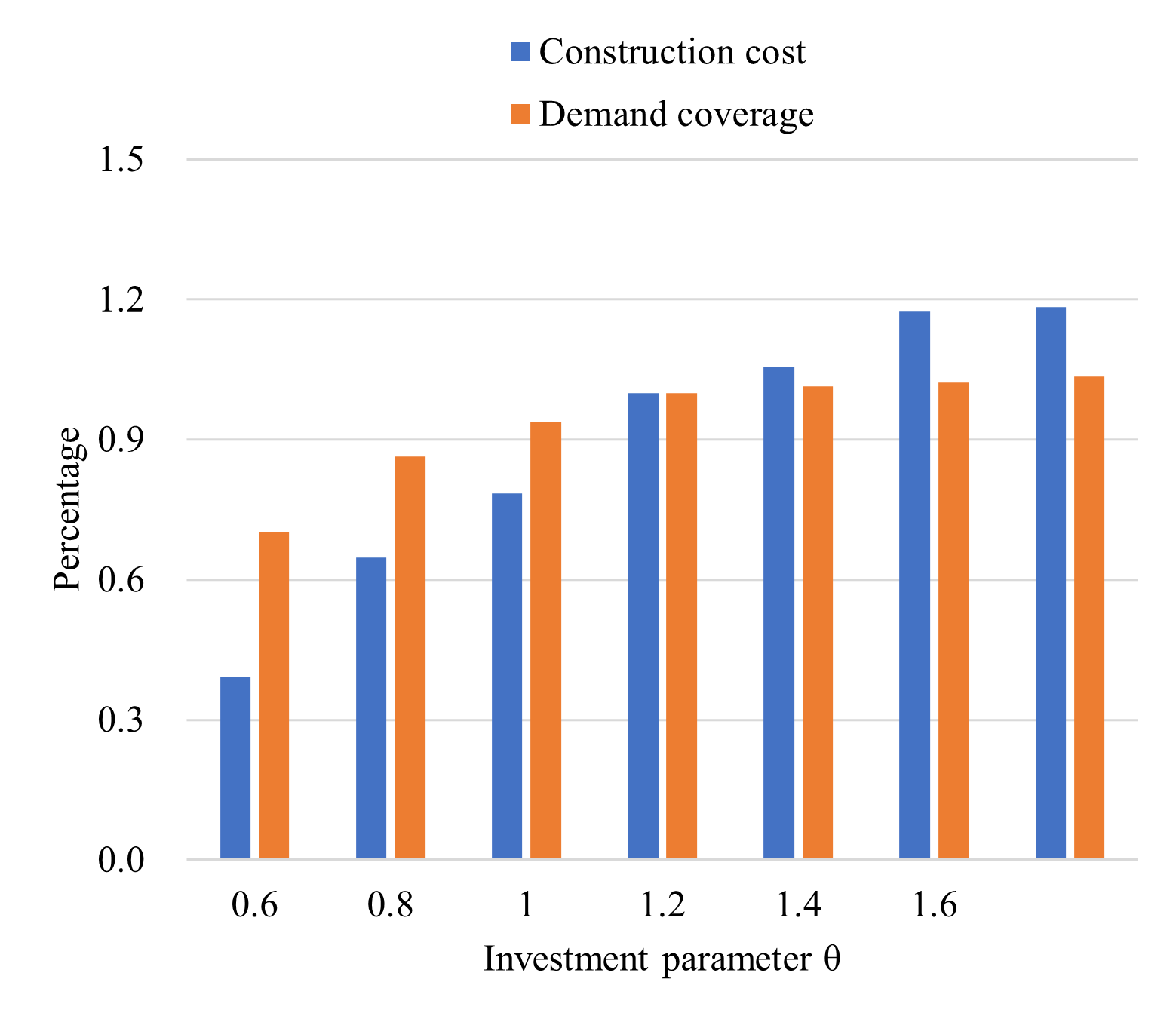}
         \caption{The total demand coverage of planning horizons}
    \label{fig:b2}
\end{figure}

In the optimization model, the cost in each horizon can not exceed an investment limit ($b=1.5$ million euro). To investigate the impact of the investment limitation, we defined an investment rate $\theta$. By setting the maximum investment to be $b * \theta $, the optimization results for $\theta =$ 0.4, 0.6, 0.8, 1.0, 1.2, 1.4, and 1.6 in each horizon were derived and the demand coverages were presented in Figure \ref{fig:b1}. The demand coverage grows with facilities building/upgrading from Horizon 1 to 5. Higher values of investment limits result in larger demand coverage in each horizon. The investment limitation has a significant impact on increasing demand coverage when $\theta$ increases from 0.4 to 1.0. It is possible that $\theta=$1.4 may leverage the full potential of investment, as the small difference between $\theta=$1.4 and $\theta=$1.6 may result from limitations on facility size and number. In addition,  the market penetration rate of freight vehicles has been set to be increased evenly across horizons, the covered demand does not change in the same manner. With the larger value of $\theta$, the slopes of lines in Figure \ref{fig:b1} become closer to the growth rate of market penetration.\\
Taking $\theta$ = 1.0 as the reference, Figure \ref{fig:b2} shows the percentages of total construction cost and demand coverage with varying $\theta$. It is noted that $\theta$ less 1.0 could produce solutions that are more cost-efficient, as the percentage of total construction cost is lower than the percentage of demand coverage. Therefore, higher investment limitation tends to increase charging demand coverage, but this effect weakens as it increases. When $\theta$ is larger than 1.0, even with higher investment, the demand coverage can not be improved significantly.


\section{CONCLUSION}
In this study, we develop a data-driven integrated framework for fast-charging facility planning to address the growing enroute freight charging demand. By analyzing highway traffic data, we extract the spatial and temporal patterns of the freight flow. The charging demand is determined based on freight traffic data and used to identify the significant traffic nodes along highways based on graph theory. We propose a candidate selection method to locate potential sites for charging stations and to-go chargers. We build a multi-period bi-objective optimization model to find optimal charging facility locations, considering minimum investment cost and maximum demand coverage. We applied the proposed framework to the Amsterdam highway network and utilized NSGA-II to solve the model. Scenario comparison reveals that the scenario (Scenario 1) considering next-horizon planning and selecting the solution with the highest demand coverage covers more charging demands when compared to other scenarios (Scenario 2 and 3). Sensitivity analysis demonstrates that higher investment can significantly increase demand coverage in each horizon ($\theta<1$), although the impact diminishes as the investment further increases. 

There are several directions for future research. We can extend the methodological framework to include logistics-related POI information, such as distribution centers and warehouses, and use the origin-destination data of electric freight vehicles. In addition, we apply Euclidean distance to select the potential locations in this study, which can be less realistic as roads in the real-world network are connected. Future research may consider obtaining the route between potential stations and freight vehicles on highways and calculating the charging demand coverage based on the real network. Finally, this study takes the highway near Amsterdam as a case study, more investigations can be done to extend the modeling problem to large highway networks.

\addtolength{\textheight}{-1cm}   



\newpage
\section*{APPENDIX}
The parameters are listed in Table \ref{tab:value}. In setting these parameters, we refer to the parameter settings in previous research \cite{Li2021, Wu2021} and have tuned the parameters based on our case study.

\begin{table}[!h]
    \centering
    \caption{Parameter settings}
    \label{tab:value}
    \begin{tabular}{ll}
    \hline
        Variable & Setting\\
        \hline
         $dist_{min}$  &  3 (km) \\
         $s$  &  5 \\
         $n $ &  5 \\               
         $c_t$ & 2000 (euro)\\
         $cap_l,l\in L$ & [0, 30, 35, 40, 45, 50] (vehicles/hour)\\
         $cap_t$ & 2 (vehicles/hour) \\         
         $p^k,k\in K$ & [0.2, 0.4, 0.6, 0.8, 1]\\
         $b^k,k\in K$ & [1.5, 1.5, 1.5, 1.5, 1.5, 1.5] (million euro) \\
         $Nmin^k,k\in K$  &  [0, 0, 0, 0, 0]\\
         $Nmax^k,k\in K$ &  [5, 5, 5, 5, 5]\\
         $Mmin^k,k\in K$  & [0, 0, 0, 0, 0] \\
         $Mmax^k,k\in K$ & [15, 15, 15, 15, 15] \\
         $c^{0,l}_s,l\in L$ & [[0, 0.5, 0.6, 0.7, 0.8, 0.9] (million euro)\\
         $c^{1,l}_s,l\in L$& [0, 0, 0.2, 0.3, 0.4, 0.5] (million euro)\\
         $c^{2,l}_s,l\in L$& [0, 0, 0, 0.25, 0.35, 0.45] (million euro)\\
         $c^{3,l}_s,l\in L$& [0, 0, 0, 0, 0.3,0.4] (million euro)\\
         $c^{4,l}_s,l\in L$& [0, 0, 0, 0, 0, 0.45] (million euro)\\
    \hline  
    \end{tabular}
\end{table}

\section*{ACKNOWLEDGMENT}
This study is supported by Shell through a research collaboration initiative with the Delft University of Technology in the space of transportation electrification.


\section*{REFERENCES}

\renewcommand{\section}[2]{}%
\bibliographystyle{unsrt}
\bibliography{ref}

\end{document}